\documentclass[twocolumn,floats,prb,aps,showpacs]{revtex4}
\usepackage[final]{graphicx}
\usepackage[dvipsnames]{color}
\usepackage{dcolumn}
\usepackage{hhline}
\DeclareGraphicsExtensions{.pdf,.eps}

\begin{document}

\title{Many-body Green's function study of coumarins for dye-sensitized solar cells}

\author{C. Faber,$^1$ I. Duchemin$^2$, T. Deutsch$^2$, X. Blase$^1$}

\affiliation{ $^1$Institut N\'{e}el, CNRS and Universit\'{e} Joseph Fourier,
B.P. 166, 38042 Grenoble Cedex 09, France. \\
$^2$INAC, SP2M/L$\_$sim, CEA Cedex 09, 38054 Grenoble, France. }

\date{\today}

\begin{abstract}
We study within the many-body Green's function $GW$ and  Bethe-Salpeter formalisms the excitation 
energies of several coumarin dyes proposed as an efficient alternative to ruthenium complexes for 
dye-sensitized solar cells. Due to their internal donor-acceptor structure, these chromophores present 
low-lying excitations showing a strong intramolecular charge-transfer character.  We show that 
combining $GW$ and Bethe-Salpeter calculations leads to charge-transfer excitation energies and
oscillator strengths in excellent agreement with reference range-separated functional studies or 
coupled-cluster calculations.  The present results confirm the ability of this family of approaches 
to describe accurately Frenkel and charge-transfer photo-excitations in both extended and finite 
size systems without any system-dependent adjustable parameter, paving the way to the study of 
dye-sensitized semiconducting surfaces. 
\end{abstract}

\maketitle


\section{Introduction}

\begin{figure}
\begin{center}
\includegraphics*[width=0.45\textwidth]{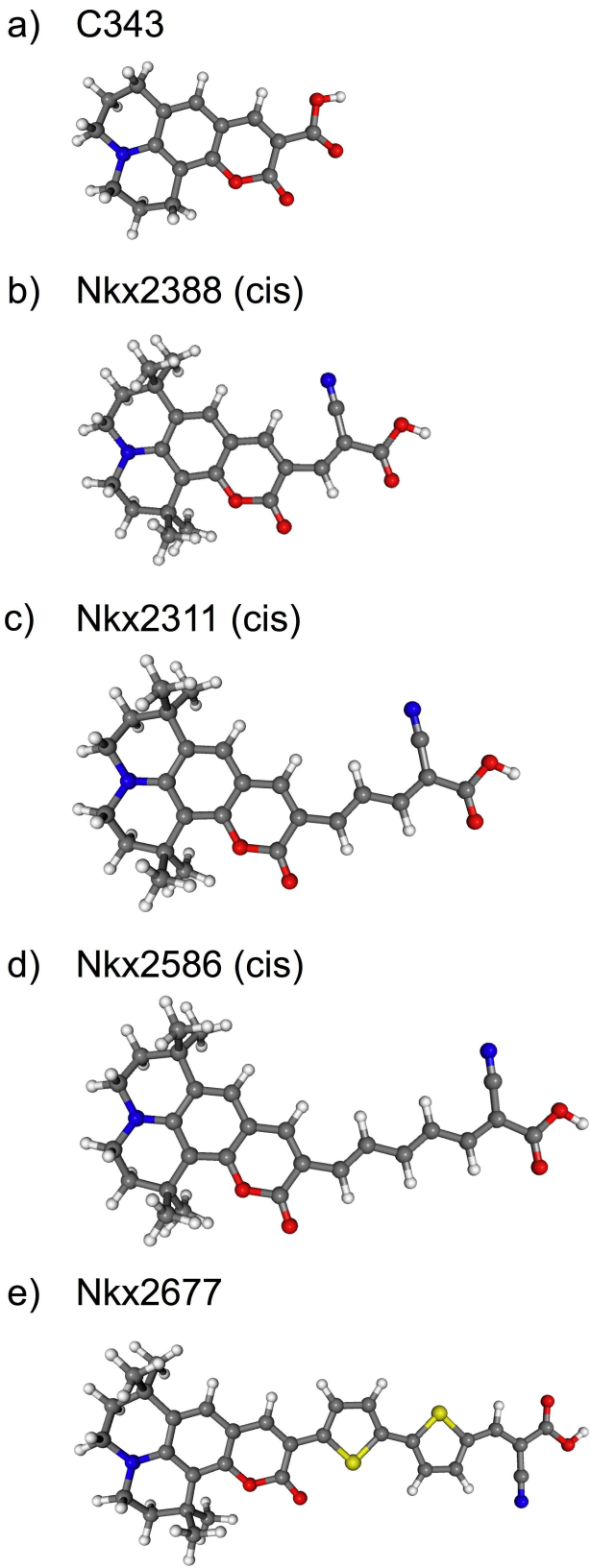}
\caption{ (Color online) Symbolic representation of the studied coumarins:
(a) parent C343, (b) NKX-2388 (cis), (c) NKX-2311 (cis), NKX-2586 (cis) and
NKX-2677. The difference with the corresponding (trans) structures is represented
in Fig.~2 for the NKX-2311 case.  Black, white, red, blue and yellow atoms represent
carbon, hydrogen, oxygen, nitrogen and sulfur, respectively. }
\label{fig1}
\end{center}
\end{figure}

Promising to become a low-cost alternative to standard silicon-based photovoltaics, dye-sensitized 
solar cells (DSSC) have been intensively studied over the past 20 years. \cite{Gratzel91,Hagfeldt10}
The most prominent modern DSSCs, the so-called Gr\"{a}tzel cells, consist of porous layers of 
titanium dioxide nanoparticles covered by molecular dyes that absorb sun light.  While most efficient
sensitizers are composed of ruthenium dye complexes, intense research is conducted so as to find  
molecular alternatives which are cheaper, easier to synthesize, and free from the resource limitations 
related to the noble metal ruthenium. As a promising direction, Hara and coworkers \cite{Hara03} 
demonstrated that coumarin-based dyes, such as the so-called NKX-2xxx family (see Fig.~\ref{fig1}),
could lead to conversion efficiencies approaching that of ruthenium-based DSSCs. 

\begin{figure}
\begin{center}
\includegraphics*[width=0.45\textwidth]{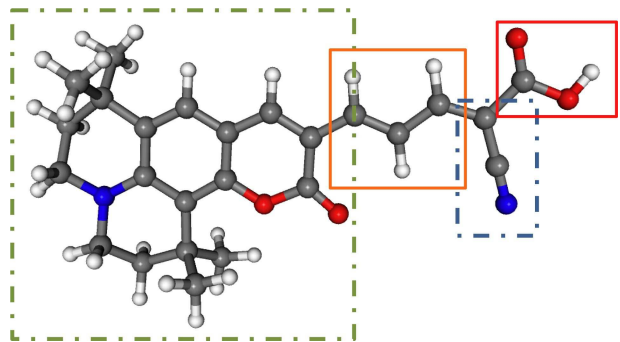}
\caption{ (Color online) Symbolic representation of the NKX-2311 (trans) coumarin dye.
This structure can be compared to the NKX-2311 (cis) structure in Fig.~1(c) which differs
from the orientation of the $\pi$ conjugated polymethine (-HC=CH-) bridge (orange frame),
the cyano (-C=N) (blue dot-dashed box) and the carboxylic (-COOH) (full red box) groups. 
The donor coumarin parent is on the left side in the green dot-dashed frame.  
The anchoring group to the semiconducting TiO$_2$ surface is the carboxylic unit.}
\label{fig2}
\end{center}
\end{figure}

Starting from the originally tested C343 coumarin \cite{Rhem96} (Fig.~\ref{fig1}a), 
the introduction of (-C=C-) methine fragments between the coumarin unit and the terminal 
(-COOH) carboxyl group (Fig.~\ref{fig2}) induces a red shift of the absorption spectrum 
with improved light harvesting in the visible range. \cite{Hara03} The resulting molecular
structures are represented in Figs.~1(b-d) showing the so-called (cis) conformations, with
(trans) structural isomers represented in Fig.~2 for one of them.
Further, inclusion of the cyano (-C=N) group enhances the acceptor 
character of the combined (-COOH) and (-C=N) cyanoacrylic acid group, increasing the 
charge-transfer (CT) character of the internal excitations, a factor believed to favor the
 injection of the photoelectron to the inorganic structure through the anchoring (-COOH) 
carboxylic unit. \cite{Hara03} Finally, the replacement of the methine spacer by thiophene 
chains \cite{Hara03b} reduces the adverse aggregation of dyes onto the TiO$_2$ surface, 
leading to the NKX-2677 dye (Fig.~1e) with a solar-energy-to-electricity conversion 
efficiency of 7.7$\%$.

Due to its donor-acceptor structure, this family of molecules became also a benchmark
for theoretical studies aiming at solving the well-known problem of describing CT
excitations within time-dependent density functional theory (TDDFT). \cite{tddft,Dreuw04}
Such difficulties arise from the lack of long-range electron-hole interaction when (semi)local
kernels are used, as a signature of the missing non-local (screened) exchange contribution.
\cite{Botti04,imagecharge}
In particular, TDDFT calculations with local, \cite{Armas12} hybrid \cite{Kurashige07,Wong08} and 
range-separated hybrid (RSH) \cite{Kurashige07,Wong08,Stein09JCP} kernels were conducted and 
compared to reference quantum chemistry coupled-cluster CC2 calculations \cite{Kurashige07} in 
order to assess the accuracy of the various approaches. The central role of CT
excitations in organic and hybrid solar cells \cite{photovoltaics} certainly explains such
efforts to develop computational techniques able to describe both Frenkel and CT excitations 
in a large variety of finite and extended systems with various screening environments.

In a few recent studies, many-body Green's function techniques, namely the so-called  $GW$ 
\cite{Hedin65,Strinati80,Hybertsen86,Godby88,Strinati88,Aulbur99,Onida02} and Bethe-Salpeter (BSE)
\cite{Sham66,Strinati82,Strinati88,Rohlfing98,Benedict98,Albrecht98,Onida02} formalisms, have 
been used to investigate the optical excitations in a family of small donor/acceptor organic dyads 
\cite{Lastra11,Blase11b,Baumeier12} for which gas phase experiments \cite{Hanazaki72} are
available. The low-lying CT excitations were found to come within 0.1 eV as compared to experiment. 
\cite{Blase11b,Baumeier12} Besides these systems for which experimental data exist, 
optical  excitations in a paradigmatic fullerene/thiophene derivative acceptor/donor 
complex, \cite{Baumeier12} and intramolecular CT excitations in a model dipeptide,
\cite{Rocca10} have also been studied within the BSE formalism. This family of Green's 
function many-body techniques has been applied since the mid-eighties 
\cite{Hybertsen86,Godby88,Aulbur99} at the \textit{ab initio} level
to study extended semiconductors and metals, but its applicability to the crucial problem of 
CT excitations in organic systems remains in its infancy and extensive tests and studies are 
still needed to rationalize its merits and limitations. 

In the present work, we analyze the (singlet) excitation energies of various coumarin-based 
molecules such as the parent C343 dye and the related NKX-2388,  NKX-2311, NKX-2586, and
NKX-2677 structures for which reference calculations are available.  We demonstrate that 
the $GW$/BSE formalism yields the low-lying charge-transfer excitation energies in excellent 
agreement with optimized (parametrized) range-separated hybrid functionals and quantum chemistry 
CC2 coupled-cluster calculations with a mean absolute error smaller than 0.06 eV. Similarly,
the associated oscillator strengths are found to agree very well with ajusted long-range 
corrected functional calculations. These results consolidate the few recent evidences 
showing that $GW$/BSE calculations can describe charge-transfer excitations in gas phase 
organic systems with an accuracy of the order of a tenth of an eV without any adjustable 
parameter. 

\section{Technical details}

Our calculations are performed with the  {\sc{Fiesta}} package \cite{Blase11b,Blase11a,Faber11a,Faber11b} 
which exploits atom-centered auxiliary  Gaussian-basis for developing the needed two-body operators such 
as the dynamical ($\omega$-frequency dependent) susceptibilities, the screened Coulomb potential 
$W({\bf r},{\bf r}';\omega)$, and the self-energy in the $GW$ approximation:

$$
 \Sigma^{GW}({\bf r},{\bf r}';\omega) = {i \over 2\pi } \int d{\omega}'  e^{i {\omega}' 0^+}
      G({\bf r},{\bf r}';\omega + {\omega}') W({\bf r},{\bf r}';{\omega}'),
$$

\noindent
where $G$ is the time-ordered one-particle Green's function. The self-energy is non-local in space 
and time (energy-dependent) and accounts for exchange and correlation in the present formalism.
Our auxiliary basis contains six \textit{exp}$(-\alpha {\bf r}^2)$ Gaussians for the radial part of
each (\textit{s}, \textit{p}, \textit{d})-channel, with an even tempered \cite{Cherkes09} distribution 
of the localization coefficients $\alpha$ ranging from 0.1 to 3.2 a.u. except for hydrogen where the range 
is set to  0.1 a.u. to 1.5 a.u. As such, our auxiliary basis contains typically 54 orbitals per atom. 
Such a basis derives from previous studies \cite{Blase11b,Blase11a,Faber11a} but with additional diffuse 
orbitals. Convergency tests can be found in Ref.~\onlinecite{Blase11a} with more details about the present 
$GW$ implementation which is similar to that of Rohlfing and coworkers. \cite{Rohlfing95,Ma10} 
We however go beyond the plasmon-pole approximation for obtaining the dynamical correlations by
using contour deformation techniques. \cite{Godby88,Blase11a,Farid99}

The single-particle states needed to build the starting screened-Coulomb potential $W$ and
Green's function $G$ are the Kohn-Sham DFT eigenstates in the local density approximation (LDA)
as provided by the {\sc{Siesta}} 
package \cite{siesta} with a large triple-zeta plus double polarization basis (TZDP) for
the valence orbitals. \cite{pseudo} Using such running parameters, comparison with $GW$/BSE 
calculations performed with a reference planewave code \cite{Yambo} showed excellent agreement 
in the case of acene/tetracyanoethylene (TCNE) donor/acceptor complexes, \cite{Blase11b} 
demonstrating the accuracy of the present Gaussian-based formalism. For sake of illustration, 
the Kohn-Sham and auxiliary basis associated with the NKX-2677 molecule (67 atoms) contain 
1110 and 3618 orbitals, respectively. 
Our structures are relaxed at the all-electron B3LYP 6-311G(d,p) level. \cite{gaussian}

We start with a $GW$ calculation designed to obtain the correct quasiparticle spectrum, 
namely the correct occupied and unoccupied single-particle energy levels.  As observed in 
several studies on gas phase (isolated) molecular systems, 
\cite{Blase11a,Faber11a,Hahn05,Tiago08,Sahar12,Abramson12}
the standard ``single-shot" $G_0W_0$(LDA) scheme tends to produce too small energy gaps
between the highest occupied (HOMO) and the lowest unoccupied (LUMO) molecular orbitals.
In the case of CT excitations with a large weight on the HOMO-LUMO transition, 
this has been shown to result in $G_0W_0$/BSE excitation energies too small as compared 
to experiment, with a mean absolute error of about 0.7 eV in the case of the small 
TCNE/acenes complexes. \cite{Blase11b} Such a problem has been largely cured by several 
groups \cite{Blase11b,Baumeier12,Blase11a,Tiago08,Sahar12,Abramson12} thanks to a  partial 
self-consistent cycle where the quasiparticle energies are reinjected to build updated 
time-ordered Green's function $G$ and screened Coulomb potential $W$, while keeping the 
eigenstates (Kohn-Sham one particle orbitals) frozen. \cite{Shishkin07} This simple self-consistent 
loop allows to obtain much better quasiparticle  \cite{Blase11a,Faber11a,Sahar12,Abramson12}
and excitation \cite{Blase11b,Baumeier12} energies as compared to experiment. Further,
the observed sensitivity of the $G_0W_0$ results to the choice of the starting zeroth-order 
eigenstates \cite{Blase11a,Blase11b,Marom11,Sahar12,Abramson12,Shishkin07,Korzdorfer12}
is significantly reduced. \cite{HFvsLDA} It is such an approach that we adopt
in the present study.

In a second step, the neutral (optical) excitation spectrum is constructed by using the 
Bethe-Salpeter equations (BSE) \cite{Sham66,Strinati82,Rohlfing98,Benedict98,Albrecht98}
that account for the electron-hole (excitonic) interactions. Deriving from Green's 
function many-body perturbation theory, the Bethe-Salpeter Hamiltonian $H^{BSE}$ is most 
commonly written in the two-body product basis $\phi_i({\bf r})\phi_j({\bf r}')$ of
the Kohn-Sham orbitals, \cite{realwfn} resulting in the following generalized eigenvalue 
problem:

\begin{eqnarray*}
 \left(
   \begin{array}{cc}
      A_{vc,v'c'}   &  C_{vc,c'v'}  \\
      -C^*_{cv,v'c'}   &  -A^*_{cv,c'v'}  \\
   \end{array}
 \right)
  \left(
   \begin{array}{c}
      X_{v'c'}    \\
      Y_{c'v'}   \\
   \end{array}
  \right)
= \Omega
  \left(
   \begin{array}{c}
      X_{vc}    \\
      Y_{cv}   \\
   \end{array}
  \right)
\end{eqnarray*}

\noindent The first diagonal block:  $\; A^{BSE} = ( H^{diag} + H^{direct} + H^{exch} )$, accounts 
for resonant transitions from occupied (indexes v,v') to unoccupied (indexes c,c') eigenstates, with 
the following contributions:

\begin{eqnarray*}
    H^{diag}_{vc,v'c'} &=& (\varepsilon_c^{GW} - \varepsilon_v^{GW}) \delta_{vv'}  \delta_{cc'},
\\
    H^{direct}_{vc,v'c'} &=& -  \int d{\textbf r} d{\textbf r'} \phi_c({\textbf r}) \phi_v({\textbf r'})
    W({\textbf r},{\textbf r'}) \phi_{c'}({\textbf r}) \phi_{v'}({\textbf r'}),
\\
   H^{exch}_{vc,v'c'} &=&  2 \int d{\textbf r} d{\textbf r'} \phi_c({\textbf r}) \phi_v({\textbf r})
    v({\textbf r},{\textbf r'}) \phi_{c'}({\textbf r}') \phi_{v'}({\textbf r}').
\end{eqnarray*}

\noindent  The second diagonal block
$(-A^*)$ accounts for non-resonant transitions from unoccupied to occupied levels. The off-diagonal 
C term couples resonant and non-resonant transitions, with C also composed of a direct and an 
exchange term which can be obtained from their A-block  analogs by switching the
(v',c') indexes. While all empty states are included in the construction of the independent-electron
susceptibilities and Green's function needed for the $GW$ calculations, the number of empty states 
included in the BSE Hamiltonian has been set to 160. 
Convergency tests for the largest NKX-2677 molecule indicate that increasing this number to 
200 decreases the excitation energy by less than 10 meV.

In what follows, we go beyond the Tamm-Dancoff approximation (TDA) by mixing resonant and 
anti-resonant contributions. Namely, we do calculate the coupling (C) and $(-C^*)$ blocks.
As shown in recent studies, \cite{Ma10,Gruning09} the TDA tends to overestimate by a few 
tenths  of an eV the excitations energies in small size systems where single-particle and
collective excitations can strongly mix. This is what we observe here with the low-lying 
excitation energies which are blue-shifted by up to 0.25 eV for the C343 dye, and 0.18 eV 
in the NKX-2677 case, when the off-diagonal coupling blocks are neglected.

\begin{figure}
\begin{center}
\includegraphics*[width=0.45\textwidth]{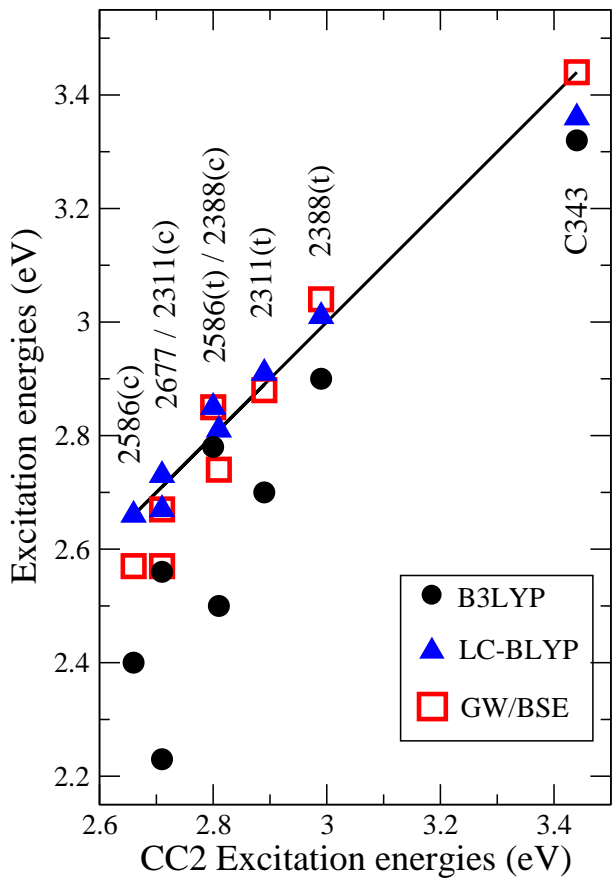}
\caption{ (Color online) Calculated lowest singlet excitation energies (in eV) as a function of the 
coupled-cluster CC2 reference values. Results in perfect agreement with the CC2 calculations should 
fall on the first diagonal (black line). The present $GW$-BSE calculations (empty red squares) 
are compared to the TD-B3LYP results of Refs.~\onlinecite{Kurashige07,Wong08} (filled black circles)
and the ``adjusted" (see text) TD-LC-BLYP data from Ref.~\onlinecite{Wong08} (blue triangles up).  
The coumarins name are indicated by their number (removing the NKX prefix) with (c) standing for -cis and 
(t) for -trans. The axes physical length is scaled according to their respective energy 
range.}
\label{fig3}
\end{center}
\end{figure}

\begin{center}
\begin{table*}
\begin{tabular}{l | m{3cm} m{2cm} m{2cm} m{3cm} m{2cm} | m{2cm} }
\hline
                 &  LDA/PBE$^{(a)}$   & B3LYP$^{(b)}$ & LC-BLYP$^{(c)}$ & BNL$^{(d)}$ J1/J2 &   CC2$^{(b)}$ &  $GW$-BSE  \\
\hline
 C343            & 2.96/3.0 (0.36/)   & 3.32 (0.60)   & 3.36 (0.57)     & 3.5/3.4 (0.7/0.6) &   3.44 (0.74) &  3.44 (0.57)  \\
NKX-2388 s-trans &                    & 2.90 (0.94)   & 3.01 (0.88)     & 3.1/2.9 (1.0/0.9) &   2.99 (1.06) &  3.04 (0.88)  \\
NKX-2388 s-cis   &                    & 2.78 (0.87)   & 2.85 (0.80)     & 2.9/2.8 (0.9/0.9) &   2.80 (1.00) &  2.85 (0.80)  \\
NKX-2311 s-trans &                    & 2.70 (1.35)   & 2.91 (1.34)     & 2.9/2.8 (1.6/1.5) &   2.89 (1.51) &  2.88 (1.37) \\
NKX-2311 s-cis   & 2.35/2.35 (1.05/)  & 2.56 (1.19)   & 2.73 (1.12)     & 2.7/2.6 (1.3/1.2) &   2.71 (1.33) &  2.67 (1.13) \\
NKX-2586 s-trans &                    & 2.50 (1.71)   & 2.81 (1.83)     & 2.8/2.6 (2.1/2.0) &   2.81 (2.01) &  2.74 (1.88) \\
NKX-2586 s-cis   & 2.10/2.15 (1.23/)  & 2.40 (1.55)   & 2.66 (1.52)     & 2.6/2.5 (1.7/1.7) &   2.66 (1.74) &  2.57 (1.59) \\
 NKX-2677        &                    & 2.23 (1.49)   & 2.67 (1.76)     & 2.7/2.5 (2.0/1.8) &   2.71 (2.17) &  2.56 (1.69) \\
\hline
\end{tabular}
\caption{Calculated lowest ($S_0 \rightarrow S_1$) singlet transition energies (in eV). The $GW$-BSE results 
calculated in the present study are compared to the TD-LDA, TD-PBE,  TD-B3LYP, TD-LC-BLYP, TD-BNL and CC2 calculations
from Refs.~\onlinecite{Armas12,Kurashige07,Wong08,Stein09JCP}. We provide the B3LYP results from
Ref.~\onlinecite{Kurashige07} which are in excellent agreement with the results of Ref.~\onlinecite{Wong08}.
The numbers in parenthesis indicate the associated oscillator strengths. \\
$^a$Ref.~\onlinecite{Armas12}. \\
$^b$Ref.~\onlinecite{Kurashige07}. \\
$^c$Ref.~\onlinecite{Wong08}. \\
$^d$Ref.~\onlinecite{Stein09JCP}.  }
\label{tablebse}
\end{table*}
\end{center}

\begin{figure}
\begin{center}
\includegraphics*[width=0.45\textwidth]{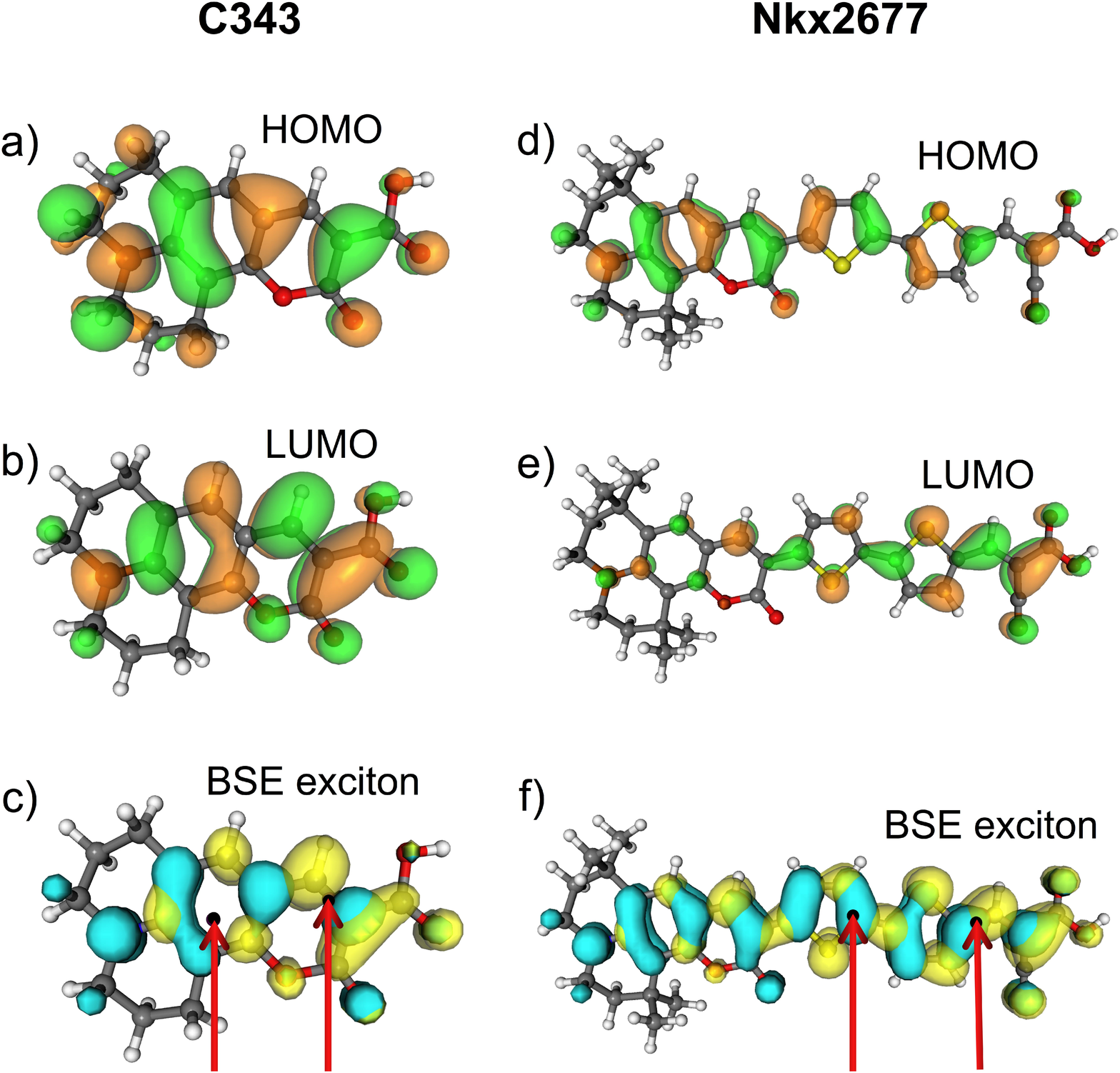}
\caption{ (Color online) (a) and (b) Isocontour representation of the C343 HOMO and LUMO Kohn-Sham
eigenstates. Different colors indicate different signs of the wavefunction. In (d) and (e), similar
plots for NKX-2677. In (c) and (f), isocontour representation of the electron (yellow) and hole
(light blue) probability distribution for the lowest C343 and NKX-2677 singlet excited states,  
respectively, as obtained within BSE. The red arrows indicate the average hole (left arrow) and 
electron (right arrow) positions. The C343 and NKX-2677 molecules are not represented on the 
same scale. }
\label{fig4}
\end{center}
\end{figure}

\section{Results and analysis}

Our results are compiled in Table~\ref{tablebse} and in Fig.~\ref{fig3} with a comparison to previously
published TDDFT and CC2 calculations. As expected, the TD-LDA and TD-PBE values from Ref.~\onlinecite{Armas12}
evidence an underestimation of the transition energies related to the CT character of the
excitations. The mean absolute error (MAE) averaged over the available data points amounts to 0.47 eV
and 0.44 eV as compared to CC2 within LDA and PBE, respectively. Further, the oscillator strength seems
to be significantly underestimated, in particular, surprisingly, in the case of the smaller C343 parent 
molecule expected to show the smallest CT character.

Thanks to its 20$\%$ of exact exchange, the B3LYP results of  Ref.~\onlinecite{Kurashige07} (filled black
circles in Fig.~\ref{fig3}) indicate a reduced MAE of 0.2 eV. 
However, the discrepancy can still be as large as 0.48 eV for the NKX-2677 structure.
This is certainly the signature that in the long-range charge-separation limit, the restricted amount of
exact exchange in the B3LYP functional \cite{b3lyp} is not enough to account for the correct electron-hole 
interaction. To illustrate that point, we plot in Fig.~\ref{fig4} the  Kohn-Sham  (a,d) HOMO and (b,e) LUMO 
eigenstates associated with the C343 and NKX-2677 structures. While the HOMO states are found to be rather 
delocalized, the LUMO in the NKX-2677 dye is clearly much more localized close to the electron-acceptor 
cyanoacrylic group, resulting in an enhanced CT character as compared to the C343 parent molecule.

The nature of the transitions can be better quantified by studying the electron and hole spatial localization 
in the excited states. This can be achieved by taking the expectation value of the electron/hole position 
operator $\delta({\bf r}-{\bf r}_e/{\bf r}_h)$ over the two-body $\psi({\bf r}_e,{\bf r}_h)$ BSE excitonic 
wavefunction, leading to an electron/hole probability of presence averaged over the hole/electron
position. The resulting densities are provided in Fig.~\ref{fig4}(c,f) with an isocontour representation
for the C343 and NKX-2677 molecules. These densities allow to obtain the mean electron/hole positions (see
red arrows in Fig.~\ref{fig4}) and the related average electron-hole separation distance which amounts to 
3.2~\AA\ in C343.  This clear CT character is certainly at the origin of the difficulties 
met by LDA or PBE to describe such an excitation. In the NKX-2677 case, this average distance increases
to 4.6~\AA\ as a signature of the enhanced CT character,  explaining that the B3LYP results 
significantly worsen from C343 to NKX-2677. 

We now come to the central results of the present study, namely the many-body perturbation theory data.
In contrast with the LDA, PBE or even B3LYP results, our $GW$-BSE values (empty red squares in Fig.~\ref{fig3}) 
are in much better agreement with the CC2 data points, with a mean absolute error of 0.06 eV. Such an 
agreement is remarkable accounting for the fact that the present $GW$/BSE approach does not contain any 
adjustable parameter. Concerning the longest NKX-2677 dye, which shows the largest discrepancy with CC2 
calculations, we observe  that our result falls within the values provided by the range-separated 
hybrid \cite{Savin96} (RSH) BNL functional study \cite{Stein09JCP} where two different strategies to 
optimize \textit{ab initio} (non-empirically) the range-separation parameter have been tested. \cite{BNLrecipe} 
As compared to the RSH-BNL study, our $GW$-BSE results differ by a MAE ranging from 0.04 eV to 0.07 eV, that 
is well within 0.1 eV. As emphasized in Ref.~\onlinecite{Stein09JCP}, the CC2 approach is also not free 
from approximations \cite{CC2basis} and differences of the order of 0.1 eV as compared to more accurate 
e.g. CASPT2 calculations are certainly to be expected.

Clearly, as compiled in Table~\ref{tablebse}, TDDFT calculations with the LC-BLYP functional \cite{Iikura01}
also provide excellent results, \cite{Wong08} with a 0.03 eV MAE as compared to CC2, smaller than our 
$GW$-BSE 0.06 eV MAE value. However, as emphasized in Ref.~\onlinecite{Wong08}, the range-separation 
parameter $\mu$ in the LC-BLYP study has been precisely adjusted to minimize the root mean square error 
with CC2 calculations.  The best-fit $\mu$ value for these systems ($\mu$=0.17 a.u.) is found to be 
much smaller than the original $\mu$=0.33 advocated by Iikura and coorkers \cite{Iikura01}. The strong 
dependence of the excitation energies as a function of $\mu$ indicates that the choice of the originally 
recommended $\mu$=0.33 value would lead to a significant overestimation of the transition energies (by as 
much as 0.3-0.4 eV, see Fig.~4 of Ref.~\onlinecite{Wong08}). This leads to the standard question of the 
choice and transferability of the range-separation parameter(s). One observes however that with the best-fit 
$\mu$ value, the correlation between LC-BLYP and CC2 results is very remarkable, showing that this 
class of systems can be described by a unique parameter. 

Bearing important consequences on the use of such dyes in DSSCs, our $GW$-BSE calculations confirm the
range-separated hybrid TDDFT and CC2 data leading to the conclusion that the onset of absorption is significantly
red shifted with increasing size length. This evolution is in clear contrast with the behavior of 
CT excitations in well separated gas phase donor/acceptor dyads where the exciton binding energy 
scales as the inverse distance between the two molecules, leading to an increase of the absorption energy
onset. However, contrary to well separated donor/acceptor systems, the quasiparticle HOMO-LUMO gap in 
donor/acceptor dyads connected by a conducting $\pi$-conjugated bridge does not remain constant with varying 
bridge length. This is clearly exemplified in Table~\ref{tablegaps} where the $GW$ HOMO-LUMO gap is 
found to quickly decrease from the C343 molecule to the longest NKX-2677 dye. Except for the large 
variation of the electronic affinity (EA) from the C343 parent to the NKX-2388 system, analysis of the 
$GW$ HOMO and LUMO quasiparticle energies indicates that this gap reduction stems both from a 
destabilisation of the HOMO (decrease of the ionization potential IP) and a stabilization of the 
LUMO (increase of the EA).  Subtracting the $GW$ HOMO-LUMO gap from the singlet $GW$-BSE excitation 
energy, one finds that the electron-hole (excitonic) binding energy $E_B$ decreases with donor-acceptor 
distance, but this effect is not strong enough to counterbalance the decrease of the HOMO-LUMO gap. 
This is an important feature which explains that the longest NKX-2677 dye is most efficient in 
harvesting photons in the visible range.

\begin{center}
\begin{table}
\begin{tabular}{l|m{0.3cm} m{1cm} m{1cm} m{1cm} m{1cm}  }
\hline
                 &     & IP   & EA   & Gap    & $E_B$    \\
\hline
 C343            &     & 7.21 & 0.60 &  6.61  &  3.17   \\
NKX-2388 s-trans &     & 7.12 & 1.33 &  5.79  &  2.75       \\
NKX-2388 s-cis   &     & 7.11 & 1.50 &  5.61  &  2.76   \\
NKX-2311 s-trans &     & 6.93 & 1.55 &  5.38  &  2.50    \\
NKX-2311 s-cis   &     & 6.92 & 1.71 &  5.21  &  2.54   \\
NKX-2586 s-trans &     & 6.76 & 1.70 &  5.06  &  2.32    \\
NKX-2586 s-cis   &     & 6.77 & 1.85 &  4.91  &  2.34   \\
 NKX-2677        &     & 6.52 & 1.62 &  4.90  &  2.33    \\
\hline
\end{tabular}
\caption{Calculated $GW$ ionization potential (IP), electronic affinity (EA),  
and HOMO-LUMO gap. The exciton binding energy $E_B$ is the $GW$ HOMO-LUMO gap 
minus the BSE (singlet) excitation energy. Energies are in eV.} 
\label{tablegaps}
\end{table}
\end{center}

\begin{figure}
\begin{center}
\includegraphics*[width=0.45\textwidth]{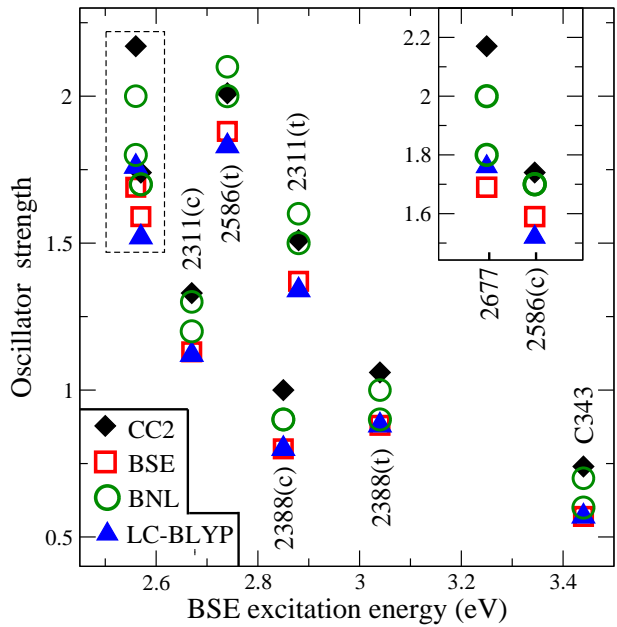}
\caption{ (Color online) Theoretical strength as a function of the $GW$-BSE excitation 
energies (in eV).  The BSE values (red empty squares) are compared to the CC2 (black 
diamond), LC-BLYP (blue triangle up), and the BNL J1/J2 (empty green circles) results. 
Results for the NKX-2677 and NKX-2586 (s-cis) in the dashed box are reproduced in the
up-right inset. }
\label{fig5}
\end{center}
\end{figure}

We close this study by commenting on the oscillator strength  associated with the calculated 
transitions (see the numbers in parenthesis in Table~\ref{tablebse} and Fig.~\ref{fig5}). Our $GW$-BSE 
values (red empty squares) lead to an excellent agreement with the parametrized 
LC-BLYP (blue up triangles) calculations, 
with a perfect match for the C343 and NKX-2388 molecules. Such an agreement is 
remarkable given the fact that the two approaches are very different. Both $GW$-BSE and LC-BLYP
provide lower values as compared to CC2 calculations, the BNL data yielding somehow intermediate
results.  The largest discrepancies occur for the NKX-2677 molecule, with a 22$\%$ difference 
between the $GW$-BSE and CC2 values.  For this system, $GW$-BSE and LC-BLYP agree again extremely 
well, within 4$\%$, while the two BNL values (see Inset) show their largest spread, indicating
that the oscillator strength for this structure is very sensitive to the chosen formalism and
related parameters.  Further analysis is needed to understand these variations from one 
type of calculation to another.  Beyond differences, one should emphasize that all studies agree 
on the fact that the NKX-2677 dye presents one of the largest oscillator strength, yet another 
factor explaining that it leads to one of the largest conversion efficiency in the coumarin family.

\section{Conclusion}

In conclusion, we have studied within the many-body Green's function $GW$ and BSE perturbation theory
the excitation energies of a family of coumarin dyes recently shown to be very promising candidates for 
replacing ruthenium-based chromophores in dye-sensitized solar cells (DSSC). In such donor-bridge-acceptor 
molecules, the lowest singlet excitations are characterized by a charge-transfer character that varies 
with the length of the $\pi$-conjugated bridge. As a result, TD-B3LYP calculations can lead to an error 
as large as $\sim$0.5 eV as compared to reference quantum chemistry coupled-cluster CC2 calculations, 
despite the 20$\%$ of exact exchange contained in its functional form. We demonstrate that the $GW$/BSE 
approach leads to an excellent agreement with CC2 data with a mean absolute error of the order of 0.06 eV
for the excitation energies. Such an accuracy is comparable to the best results provided by TDDFT 
calculations with optimized long-range corrected range-separated hybrids, but with a parameter-free 
approach that performs equally well for extended insulating or metallic systems and gas phase organic 
molecules. Such an excellent agreement is also demonstrated for the related oscillator strengths. 
The ability of the $GW$/BSE approach to describe both Frenkel and charge-transfer excitations in 
finite size molecular systems or extended semiconductors originates in particular from the use of 
the screened Coulomb potential $W$ that automatically adjusts the strength and range of the Coulomb 
interactions.  This flexibility may proove as a significant advantage in the study of DSSCs where 
both the organic dye and the extended TiO$_2$ semiconductor must be treated with sufficient accuracy. 

\textbf{Acknowledgements.}
I.D. acknowledges funding from the CEA ``Eurotalent" program and C.F. a joint CEA/CNRS BDI fellowship.  
The authors are indebted to L. Kronik for sending the initial geometries of the coumarins studied 
in this work, and to C.~Attaccalite and V.~Olevano for many suggestions and critical readings of 
our manuscript. Calculations have been performed thanks to French national supercomputing IDRIS
center at Orsay under contract n$^o$ i2012096655. 



\begin{thebibliography}{150bliography}

\bibitem{Gratzel91}
B. O'Regan and M. Gr\"{a}tzel, \textit{Nature} \textbf{353}, 737 (1991). 

\bibitem{Hagfeldt10}
A. Hagfeldt, G. Boschloo, L. Sun, L. Kloo, and H. Pettersson,
Chem. Rev.  \textbf{110}, 6595-6663 (2010).

\bibitem{Hara03}
K. Hara, T. Sato, R. Katoh, A. Furube \textit{et al.} 
J. Phys. Chem. B \textbf{107}, 597-606 (2003).

\bibitem{Rhem96}
J. M. Rhem, G. L. McLendon, Y. Nagasawa, K. Yoshihara,
J. Moser and M. Gr\"{a}tzel, J. Phys. Chem. \textbf{100}, 9577 (1996).

\bibitem{Hara03b}
K. Hara,  M. Kurashige,  Y. Dan-oh,  C. Kasada , A. Shinpo,  S. Suga,  
K. Sayama and H. Arakawa, New J. Chem. \textbf{27}, 783-785 (2003).

\bibitem{tddft}
E. Runge and E.K.U. Gross, Phys. Rev. Lett. \textbf{52}, 997 (1984);
M.A.L. Marques, C.A. Ullrich, F. Nogueira, A. Rubio, K. Burke, and E.K.U. Gross (eds.), 
\textit{Time-Dependent Density Functional Theory} (Springer-Verlag, 2006);
Mark E. Casida, J. Mol. Struct. (Theochem) \textbf{914}, 3 (2009).

\bibitem{Dreuw04}
A. Dreuw,  M. Head-Gordon, J. Am. Chem. Soc.  \textbf{126}, 4007-4016 (2004).

\bibitem{Botti04}
Similar problems arise in extended solids for Wannier excitons with large effective bohr radius
characterized by a large average electron-hole distance. See:
S. Botti, F. Sottile, N. Vast, V. Olevano, L. Reining, H.-C. Weissker, A. Rubio, G. Onida, 
R. Del Sole, and R. W. Godby, Phys. Rev. B \textbf{69}, 155112 (2004). 

\bibitem{imagecharge}
An interesting ``static" analog, illustrating the role of non-local exchange for long-range
electron-hole interaction, is the image-charge potential problem within DFT. For an analysis 
in the case of metallic surfaces, see:
I.D. White, R.W. Godby, M.M. Rieger, and R.J. Needs,
Phys. Rev. Lett. \textbf{80}, 4265  (1998) 


\bibitem{Armas12}
R. S\'{a}nchez-de-Armas, M. \'{A}ngel San Miguel, J. Oviedo and
J. Fdez. Sanz, Phys. Chem. Chem. Phys. \textbf{14}, 225 (2012).

\bibitem{Kurashige07}
Y. Kurashige, T. Nakajima, S. Kurashige, K. Hirao and  Y. Nishikitani,
J. Phys. Chem. A \textbf{111}, 5544-5548 (2007).

\bibitem{Wong08}
B.M. Wong and J.G. Cordaro, J. Chem. Phys. \textbf{129}, 214703 (2008).

\bibitem{Stein09JCP}
T. Stein, L. Kronik  and R. Baer,
J. Chem.  Phys. \textbf{131}, 244119 (2009).


\bibitem{photovoltaics}
N.S. Sariciftci, L. Smilowitz, A.J. Heeger, F. Wudl, \textit{Science} \textbf{258}, 1474 (1992); 
S.A. Jenekhe, J.A. Osaheni, \textit{Science} \textbf{265}, 765 (1994);
J.L. Bredas, D. Beljonne, V. Coropceanu, J. Cornil, Chem. Rev. \textbf{104},  4971 (2004);
Y. Kanai and  J.C. Grossman, Nano Lett. \textbf{8}, 3049 (2004).


\bibitem{Hedin65}
L. Hedin, Phys. Rev. \textbf{139}, A796 (1965).

\bibitem{Strinati80}
G. Strinati, H.J. Mattausch, W. Hanke,  Phys. Rev. Lett. \textbf{45}, 290 (1980);
\textit{ibid}, Phys. Rev. B \textbf{25}, 2867 (1982).

\bibitem{Hybertsen86}
M.S. Hybertsen, S.G. Louie, Phys. Rev. B \textbf{34}, 5390 (1986).

\bibitem{Godby88}
R.W. Godby, M. Schl\"{u}ter, and L.J. Sham, Phys. Rev. B \textbf{37}, 10159 (1988).

\bibitem{Strinati88}
G. Strinati, Rivista dek Nuovo Cimento \textbf{11}, 1-86 (1988).

\bibitem{Onida02}
G. Onida, L. Reining, A. Rubio, Rev. Mod. Phys. \textbf{74}, 601 (2002).

\bibitem{Aulbur99}
For a compilation of $GW$ results in the crystalline solid phase, see e.g.
W.G. Aulbur, L. Jonsson, J.W. Wilkins, in \textit{Solid State Physics},
edited by H. Ehrenreich (Academic, Orlando, 1999), Vol. 54, p. 1.


\bibitem{Sham66} 
L.J. Sham and T.M. Rice, Phys. Rev. \textbf{144}, 708 (1966); 
W. Hanke and L.J. Sham, Phys. Rev. Lett. \textbf{43}, 387 (1979).

\bibitem{Strinati82}
G. Strinati, Phys. Rev. Lett. \textbf{49}, 1519 (1982); 
H. J. Mattausch, W. Hanke, and G. Strinati, Phys. Rev. B \textbf{27}, 3735 (1983);
Phys. Rev. B \textbf{29}, 5718 (1984).

\bibitem{Rohlfing98}
M. Rohlfing, S.G. Louie, Phys. Rev. Lett. \textbf{80}, 3320 (1998)

\bibitem{Benedict98}
L.X. Benedict, E. Shirley, R.B. Bohn, Phys. Rev. Lett. \textbf{80}, 4514 (1998).

\bibitem{Albrecht98}
S. Albrecht, L. Reining, R. Del Sole, G. Onida, Phys. Rev. Lett. \textbf{80}, 4510 (1998).


\bibitem{Lastra11}
J.M. Garcia-Lastra, K.S. Thygesen, Phys. Rev. Lett. \textbf{106}, 187402 (2011).

\bibitem{Blase11b}
X. Blase, C. Attaccalite, Appl. Phys. Lett. \textbf{99}, 171909 (2011).

\bibitem{Baumeier12}
Bj\"{o}rn Baumeier, Denis Andrienko, and Michael Rohlfing,
J. Chem. Theory Comput., Article ASAP (2012).

\bibitem{Hanazaki72}
I.J. Hanazaki, Phys. Chem. \textbf{76}, 1982 (1972).

\bibitem{Rocca10}
D. Rocca, D.Y. Lu, and G. Galli, J. Chem. Phys. \textbf{133}, 164109 (2010).


\bibitem{Blase11a}
X. Blase, C. Attaccalite, V. Olevano, Phys. Rev. B \textbf{83}, 115103 (2011). 

\bibitem{Faber11a}
C. Faber, C. Attaccalite, V. Olevano, E. Runge, X. Blase, 
Phys. Rev. B \textbf{83}, 115123 (2011).

\bibitem{Faber11b}
C. Faber, J. Laflamme Janssen, M. C\^{o}t\'{e}, E. Runge, and X. Blase,
Phys. Rev. B \textbf{84}, 155104 (2011).


\bibitem{Cherkes09}
I. Cherkes, S. Klaiman, and N. Miseyev, Int. J. Quantum Chem. \textbf{109}, 2996 (2009);
and references therein.

\bibitem{Rohlfing95}
M. Rohlfing, P. Kr\"{u}ger, J. Pollmann, Phys. Rev. B \textbf{52}, 1905 (1995).

\bibitem{Ma10}
Y. Ma, M. Rohlfing, C. Molteni, Phys. Rev. B \textbf{24}, 241405 (2009);
\textit{ibid.}, J. Chem. Theory Comput.  \textbf{6}, 257-265 (2010);
and references therein.

\bibitem{Farid99}
B. Farid, in \textit{Electron Correlation in the Solid State}, edited by N. H.
March (World Scientific, Singapore, 1999), p. 217, and references therein.


\bibitem{siesta}
J.M. Soler, E. Artacho, J.D. Gale, A. Garc\'{i}a, J. Junquera, P. Ordej\'{o}n and D. S\'{a}nchez-Portal,
J. Phys.: Condens. Mater \textbf{14}, 2745-2779 (2002).
The {\sc{Siesta}} confinement criteria on basis orbitals is released by imposing a very large cut-off 
radius  ($\sim$ 12 a.u.)

\bibitem{pseudo}
We use standard norm-conserving pseudopotentials. See: N. Troullier and J.-L. Martins,
Phys. Rev. B \textbf{43}, 1993 (1991).

\bibitem{Yambo}
A. Marini, C. Hogan, M. Gr\"uning, D. Varsano, Comput. Phys. Commun. \textbf{180}, 1392 (2009).

\bibitem{gaussian}
M. J. Frisch \textit{et al.}, GAUSSIAN 09, Revision B.01, Gaussian, Inc., Wallingford CT 2010.


\bibitem{Hahn05}
P.H. Hahn, W.G. Schmidt and F. Bechstedt, Phys. Rev. B \textbf{72}, 245425 (2005). 
In this study, the importance of correcting the starting eigenvalues in the 
construction of $W$ is emphasized for molecular systems, bringing some light 
in the discussions concerning the update of the eigenvalues in the Green's 
function only (the $GW_0$ scheme) or in both  $G$ and $W$. See also
Ref.~\onlinecite{Shishkin07} for extended solids. 

\bibitem{Shishkin07}
Weidong Luo, Sohrab Ismail-Beigi, Marvin L. Cohen, and Steven G. Louie,
Phys. Rev. B \textbf{66}, 195215 (2002);
M. Shishkin and G. Kresse, Phys. Rev. B \textbf{75}, 235102 (2007);
and references therein.

\bibitem{Tiago08}
L. Tiago, P. R. C. Kent, R. Q. Hood, and F. A. Reboredo,
J. Chem. Phys. \textbf{129}, 084311 (2008).

\bibitem{Sahar12}
S. Sharifzadeh, A. Biller, L. Kronik, J.B. Neaton, Phys. Rev. B \textbf{85}, 125307 (2012).

\bibitem{Abramson12}
S. Refaely-Abramson, S. Sharifzadeh, N. Govind, J.B. Neaton, R. Baer, L. Kronik,
arXiv:1203.2357.

\bibitem{Marom11}
N. Marom, X. Ren, J.E. Moussa, J. R. Chelikowsky and L. Kronik,
Phys. Rev. B \textbf{84}, 195143 (2011).

\bibitem{Korzdorfer12}
T. K\"{o}rzd\"{o}rfer,  N. Marom, Phys. Rev. B \textbf{86}, 041110(R) (2012).

\bibitem{HFvsLDA}
As shown in Ref.~\onlinecite{Blase11a} for a large family of organic molecules, this partial 
self-consistency scheme leads to nearly identical quasiparticle HOMO-LUMO gaps starting either 
from DFT-LDA or HF-like eigenvalues, as a clear signature of the reduced sensitivity to the
starting mean-field approximation. The $G_0W_0$(HF) results were shown to be much closer to 
the self-consistent $GW$ results than $G_0W_0$(LDA).  Similar trends have been observed 
recently with fully self-consistent $GW$ calculations on isolated atoms or small molecules 
for which such calculations are feasible.  \cite{Roostgard10,Ke11,Bruneval12} This conclusion
is expected to hold only for small systems with reduced screening. For hybrid systems such
as Gr\"{a}tzel cells, with molecular dyes deposited on semiconducting surfaces, neither the
$G_0W_0$(LDA) nor the $G_0W_0$(HF) approximation are expected to be reliable for both  the
donor (dye) and the acceptor (Ti$O_2$), emphasizing the importance of self-consistency in
some form, or the use of hybrid functionals \cite{Marom11,Korzdorfer12} that may perform
well for both systems.

\bibitem{Roostgard10}
C. Rostgaard, K. W. Jacobsen, and K. S. Thygesen,
Phys. Rev. B \textbf{81}, 085103  (2010). 

\bibitem{Ke11}
San-Huang Ke, Phys. Rev. B \textbf{84}, 205415  (2011). 

\bibitem{Bruneval12}
Fabien Bruneval,  J. Chem. Phys. \textbf{136}, 194107 (2012).

\bibitem{realwfn}
For isolated systems, wavefunctions can be taken to be real and we do not consider here the complex
conjugated eigenstates. 




\bibitem{Gruning09}
M. Gr\"{u}ning, A. Marini, X. Gonze, Nano Lett. \textbf{9}, 2820 (2009).

\bibitem{b3lyp}
A.D. Becke, J. Chem. Phys. \textbf{98}, 5648 (1993).

\bibitem{Savin96}
A. Savin, in \textit{Recent Developments and Applications of Modern Density
Functional Theory}, edited by J. M. Seminario bibitemElsevier, Amsterdam,
Chap. 9, pp. 327-354 (1996);
T. Leininger, H. Stoll, H. J. Werner, and A. Savin, Chem. Phys. Lett.  \textbf{275}, 151 (1997);
J. Toulouse, F. Colonna, and A. Savin, Phys. Rev. A \textbf{70}, 062505 (2004).

\bibitem{BNLrecipe} In the BNL approach, the range separation parameter  is obtained \textit{ab initio} by 
minimizing the MAE between the Kohn-Sham HOMO and/or LUMO eigenvalues and the corresponding quantities 
obtained with a much more accurate total energy difference $\Delta$SCF approach for the neutral and charged 
systems. Depending on the chosen criteria (optimization of the HOMO only of the neutral and charged systems, 
or of the HOMO and LUMO of the neutral system, the so-called ``J1" and ``J2" schemes), slightly different 
values can be obtained. See Ref.~\onlinecite{Stein09JCP}.   

\bibitem{CC2basis}
It is to be observed that due to computational costs at the time of publication, the CC2 calculations 
were performed with a limited SV(P) basis. See Ref.~\onlinecite{Kurashige07}.

\bibitem{Iikura01}
H. Iikura, T. Tsuneda, T. Yanai, and K. Hirao, J. Chem. Phys. \textbf{115}, 3540 (2001).

\end{thebibliography}
\end{document}